# Surface-Normal Free-Space Beam Projection via Slow-Light Standing Wave Resonances in Extra-Large Near-Zero Index Grating Couplers

Alexander Yulaev[1,2], Daron A. Westly[1], and Vladimir A. Aksyuk[1]*

[1]Physical Measurement Laboratory, National Institute of Standards and Technology, Gaithersburg, MD 20899, USA.

[2]Department of Chemistry and Biochemistry, University of Maryland, College Park, MD 20742, USA.

alexander.yulaev@nist.gov

daron.westly@nist.gov

*Corresponding author: Vladimir Aksyuk

Project Leader, Photonics and Plasmonics

Microsystems and Nanotechnology Division

Physical Measurement Laboratory, NIST

100 Bureau Dr., stop 6811

Gaithersburg, MD 20899-8412

Ph: (301) 975-2867

vladimir.aksyuk@nist.gov

# Abstract

On-chip grating couplers directly connect photonic circuits to free-space light. The commonly used photonic gratings have been specialized for small areas, specific intensity profiles and non-vertical beam projection. This falls short of the precise and flexible wavefront control over large beam areas needed to empower emerging integrated miniaturized optical systems that leverage volumetric light matter interactions, including trapping, cooling, and interrogation of atoms, bio- and chemi- sensing and complex free-space interconnect. The large coupler size challenges general inverse design techniques, and solutions obtained by them are often difficult to physically understand and generalize. Here by posing the problem to a carefully constrained computational inverse design algorithm capable of large area structures, we discover a qualitatively new class of grating couplers. The numerically found solutions can be understood as coupling an incident photonic slab mode to a spatially extended slow-light (near-zero refractive index) region, backed by a Bragg reflector. The structure forms a spectrally broad standing wave resonance at the target wavelength, radiating vertically into free space. A reflection-less adiabatic transition critically couples the incident photonic mode to the resonance, and the numerically optimized lower cladding provides 70 % overall theoretical conversion efficiency. We have experimentally validated efficient surface normal collimated emission of ≈ 90 μm full width at half maximum Gaussian at the thermally tunable operating wavelength of ≈ 780 nm. The variable-mesh-deformation inverse design approach scales to extra-large photonic devices, while directly implementing the fabrication constraints. The deliberate choice of smooth parametrization resulted in a novel type of solution, which is both efficient and physically comprehensible.

INTRODUCTION

The ability to project arbitrarily shaped, wide, collimated beams directly from a photonic chip to free space is key to realizing miniaturized chip-scale devices for optical spectroscopy[1], atom interrogation and trapping[2-6], chemi- and bio- sensing[7-12], and optical communications[13]. To achieve the desired free-space light-matter interaction, a photonic device has to efficiently couple to free-space beams with well-defined polarization, intensity and phase profiles propagating over millimeter-long distances. Beams projected normally to the photonic chip surface are particularly desired for a range of applications to simplify optical alignment and integration with other planar components and to reduce packaging costs.

Small-scale gratings are widely employed in photonics to convert sub-micrometer wide waveguide modes to free-space radiation[14-32], primarily for fiber coupling. The major limitation of conventional grating couplers is their limited ability to match photonics to spatially extended free-space modes, typically covering only ≈ 100× in mode field diameter expansion. The typical narrow photonically emitted free-space beams have a relatively short Rayleigh range of ≈ 100 μm, and their fast divergence due to diffraction makes them unsuitable for implementing hybrid 3D microsystems, where light-matter interaction occurs in volume rather than at surfaces.

Exciting recent developments in designing large photonic platforms such as a 300 μm long extreme mode converter[33], metasurface-integrated grating couplers[34,35], weakly scattering gratings in thin $SiN_x$ layers[36], and the grating Segmented Planar Imaging Detector for Electro-optical Reconnaissance (SPIDER) telescope[37] open up large sizes with control over intensity, phase and polarization state. They include reciprocal coupling of wide collimated or high-numerical-aperture focusing free-space beams with well-controlled properties to photonic modes for trapping and interrogating chip-scale atomic systems and realizing interferometric imaging systems based on photonic integrated circuits.

However, vertical emission of spatially broad free-space beams remains particularly inefficient due to the intrinsic symmetry of vertically etched grating structures. This difficulty is apparent in the reciprocal problem setting, where an incident surface-normal free-space beam is evenly coupled to the left- and right-propagating grating modes in the locally left-right symmetric grating.

Dispersion engineering based on slow-light in plasmonic and photonic systems, which is typically used to realize time domain processing, ultra-compact optical buffers, optical delay lines, and all-optical signal processors[38-40] can help mitigate both a strong chromatic dispersion and significant slab mode back reflection while maintaining the surface-normal light projection. Moreover, lowering and controlling a waveguide group velocity is a practical tool for modulating light intensity. Plasmonic waveguides, quasi-periodic plasmonic nanostructures and their hybrids with photonic waveguides have been proposed to slow down or even stop light waves[41-43]. Unfortunately, plasmonic systems are too intrinsically lossy to be practically implemented. Near-zero index (NZI) photonics – another class of structured dielectrics – enabling the static light due to decoupling of spatial and temporal fields, exhibit exotic phenomena such as light tunneling, perfect phase matching, light trapping, highly directed emission, and enhanced nonlinearities that can also facilitate tailoring of light properties[44-47].

Here, we utilize a deformation-based inverse design method – a recently introduced qualitatively different approach for solving the problem – to devise a 300 μm long photonic grating capable of projecting a 90 μm full width at half maximum (FWHM), collimated free-space beam exactly normally to the surface. Distinct from the inverse-design algorithms relying on optimizing the material properties in a design region, we make use of spatially varying induced deformations applied to a uniform grating, which is essential to handle photonic structures spanning over 600 waves. The obtained solution contains a structured slow-light NZI region supporting spatially broad resonances, including the single-peak fundamental half-wave and the higher-order standing waves, observed by their characteristic free-space radiation patterns. The input light is critically coupled to the resonances via a reflection-free adiabatic transition designed by the algorithm, achieving full impedance matching between the photonic and the free-space radiation mode. Our photonic device bridges the $10^5$ area mismatch between the photonic waveguide and surface-normal free-space radiation with 70 % theoretical conversion efficiency. We have experimentally confirmed efficient, surface-normal collimated emission, where each mode can be readily selected by tuning either the laser wavelength or the device temperature. These results demonstrate the ability of intelligently parametrized inverse design to find novel classes of solutions that are physically comprehensible and, therefore, can be generalized and adapted for different specific application requirements, e.g. different mode shapes and wavelength ranges. The developed surface-normal photonic emitter design is readily scalable for visible, telecom, and ultraviolet wavelength ranges to enable new applications in integrated photonics.

RESULTS

*Design of the resonant grating*

We follow the commonly used approach for coupling of a highly confined waveguide mode to a surface-normal wide collimated radiation by first converting the waveguide mode into a wide, collimated, vertically confined slab mode, followed by outcoupling of the slab mode to free space. The choice of $Si_3N_4$ as a waveguide core material both provides high refractive index contrast and covers a broad spectral range, including visible wavelengths, at low losses. As illustrated in Figure 1 a, for the first stage we use a photonic evanescent wave expander with a precisely designed gap between a single-mode waveguide and a slab, as reported previously[33,34,48]. The focus of this manuscript is the second stage, which uses the numerically optimized grating supporting a spatially broad standing wave resonance for coupling the slab mode to free space.

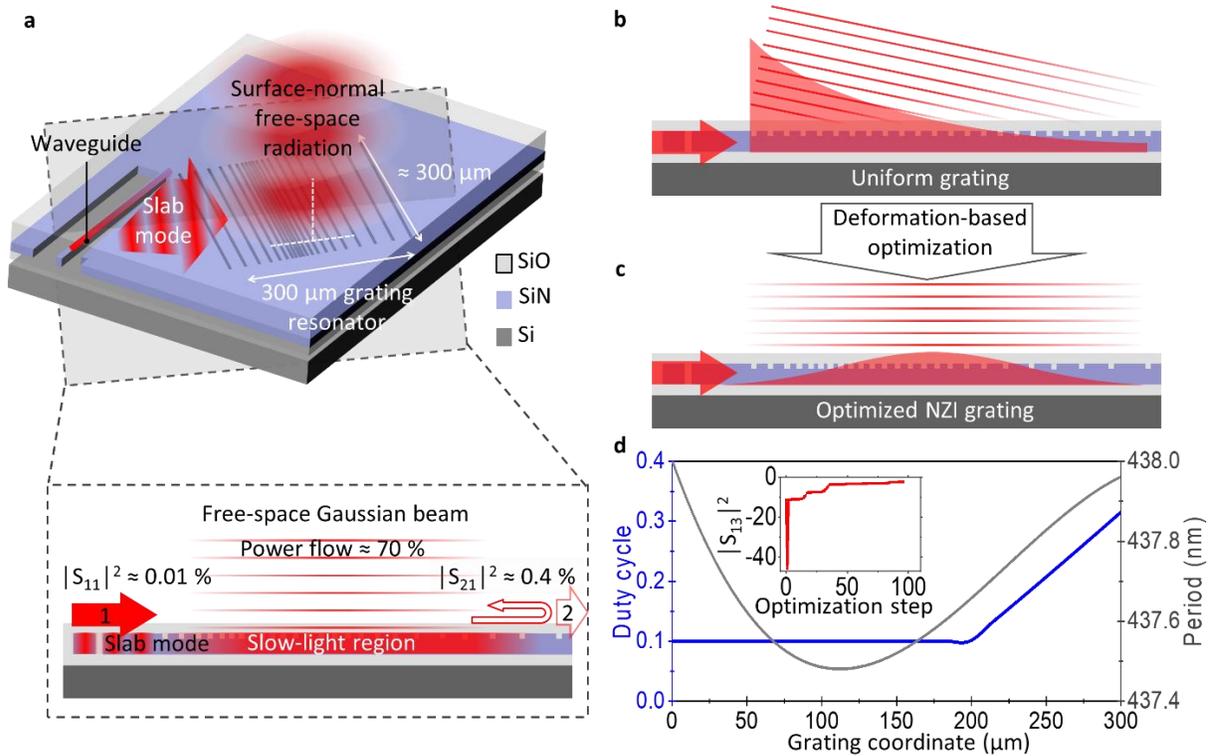

**Fig. 1 Resonant photonic grating design. a**, 3D Schematic illustrating the principal of operation. The dash box depicts the cross-sectional cartoon of the resonant grating performance. First, a waveguide single mode expands across the evanescent gap forming a collimated slab wave. Next, the apodised grating projects a surface-normal wide collimated beam into free-space from a slab mode. Arrows with 1 and 2 point out the incident and transmitted slab waves, respectively. **b** and **c** show a uniform and an apodised NZI grating, respectively. The latter is optimized by deforming the uniform grating, spatially varying the period and DC to maximize slab-to-free-space coupling. **d**, The optimized DC (blue) and period (gray) along the grating found using the inverse-design FEM. The inset depicts the optimization evolution: $|S_{13}|^2$ i.e. slab-to-free-space coupling efficiency in dB units relative to unity vs. optimization iteration.

Our solution strategy relies on an automated inverse design – a fast and general approach of searching for optical structures that optimize functional performance – which has recently revolutionized the computational photonic methods[49,50]. To obtain the grating structure, we use a gradient-based adjoint optimization algorithm combined with the finite element method (FEM). We set a cost function to maximize the coupling efficiency from a transverse electric (TE)-polarized photonic wave to a surface-normal, 100 μm waist, collimated free-space Gaussian beam. Instead of spatially varying material properties, we induce smooth deformations of an initially uniform period grating[33]. The deformation is implemented by a continuous motion of mesh elements, resulting in the continuous positioning of the grating groove walls with arbitrary resolution independently of the mesh size and without re-meshing. Owing to the wide and collimated input (slab) and output (free-space) modes, the grating modeling is reduced from a 3D to a computationally simpler 2D TE wave scattering problem. Another advantage of the deformation-based method is a binary materials choice throughout the optimization, which does not require forcing intermediate solutions of the continuously varying materials properties to satisfy the binary material constraint.

Any arbitrary grating geometry can be achieved, topologically, by continuously deforming an initially uniform periodic grating. Given the slowly varying Gaussian function as the desired free-space beam profile, we further restrict the deformations to the grating period and duty cycle (DC) expressed by independent 3rd and 4th order polynomials with coefficients that serve as the control parameters. Here the grating is deformed globally according to the varying period, and then each period is further deformed locally to achieve the desired spatially varying duty cycle. This parametrization restricts the solutions to smoothly varying duty cycles and periods. While this solution space effectively excludes more exotic situations, such as having more than a single groove in any given period, this is a deliberate choice aimed at obtaining optimized solutions that are easier to physically understand. During inverse-design optimization, we allow both the groove depth and bottom $SiO_2$ cladding thickness to vary via vertical deformation, while keeping each of them uniform across the device. The substrate is the only object breaking the up-down symmetry in the problem. It is known that efficiency can be improved further by using more complex up-down asymmetric designs, such as a double-side grating[51]. However, such designs are more complex to fabricate and, while within the scope of our inverse-design approach, they are not considered in this work. Due to practical nanofabrication limits, we constrain the minimal DC to 0.1.

Starting from the uniform duty cycle and period (Figure 1 b), the inverse design algorithm explores the space of parametrized gradual deformations without further human intervention, and finds an optimized grating geometry (Figure 1 c) that emanates an ≈ 118 μm FWHM surface-normal collimated beam (the red curves in Figures 2 a and b). The resulting DC and period profiles can be qualitatively understood by noting that at the grating coordinate $x \approx 100$ μm the grating's stop-band edge is just below the frequency of the incident light. First, to minimize reflection and maximize coupling of the slab wave into the photonic structure, the grating parameters vary smoothly on the input to provide full impedance matching between the slab mode and the grating (Figure 1 d). The period adiabatically changes within the sub-nanometer range from the $x = 0$ μm where a slab wave enters the grating to the middle of the device, decelerating the leaky grating mode due to the bandgap edge proximity and hence increasing the light intensity. Next, both the DC and the period ramp at the opposite end to reflect the remaining grating power backward. This forms a slow-light standing wave with a single half-period of ≈ 200 µm at the optimization wavelength. The grating mode at the top branch of the band diagram is a leaky mode, efficiently coupling to free-space, and the standing wave nature ensures surface-normal radiation.

Overall, the optimized photonic structure provides the critical coupling of the slab wave to the radiating resonant grating, resulting in only ≈ 0.01 % power slab wave reflection from the grating, ≈ 0.4 % power slab wave transmission, and ≈ 70 % conversion to the upward-pointing free-space beam (Figure 1 a) numerically modeled at the optimization wavelength. Because the slab mode in a thin, partially etched grating layer radiates almost equal powers up and down and given only partial reflectivity of the underlying oxide-silicon interface, optical losses into the Si substrate cannot be fully eliminated by only optimizing the lower oxide cladding thickness.

By numerically examining the device transmission spectrum to free space, we see the bandgap, the fundamental resonance (quality factor of ≈ 3000) and other standing-wave resonances at blue detuned wavelengths, separated by regions of reduced emission, as shown in Figure 2. Although the photonic emitter is inverse-designed to maximize only the conversion efficiency at a single wavelength of the fundamental mode (780 nm), the other resonances demonstrate peak outcoupling power values that are lower but comparable to that of the fundamental (Figure 2 d). The resonances' field and intensity profiles show integer numbers of half-periods along the grating (Figures 2 a,b), while their standing wave nature is confirmed by the nearly flat phase profile for the fundamental and the piece-wise flat phase profiles for the higher order resonances in Figure 2 c. Hence the apodised grating operates in an NZI regime, where fields' spatial distribution decouples from the temporal field variation and acquires a static-like character[46]. All the modes are collimated and outcoupled close to the chip's surface-normal, demonstrating near-zero angle refraction typical for NZI systems[45] (Figure 2 c, Figure S1). Figure 2 e depicts the angular dependence of the far-field free-space electric field (solid curve) and the Fourier transformed Gaussian fit of the fundamental mode (dotted curve) assuming a perfectly collimated surface-normal beam projection. From the far-field analysis, we infer that the free-space beam coupled out from the fundamental grating standing wave is diffraction limited and emanating very close to a surface normal.

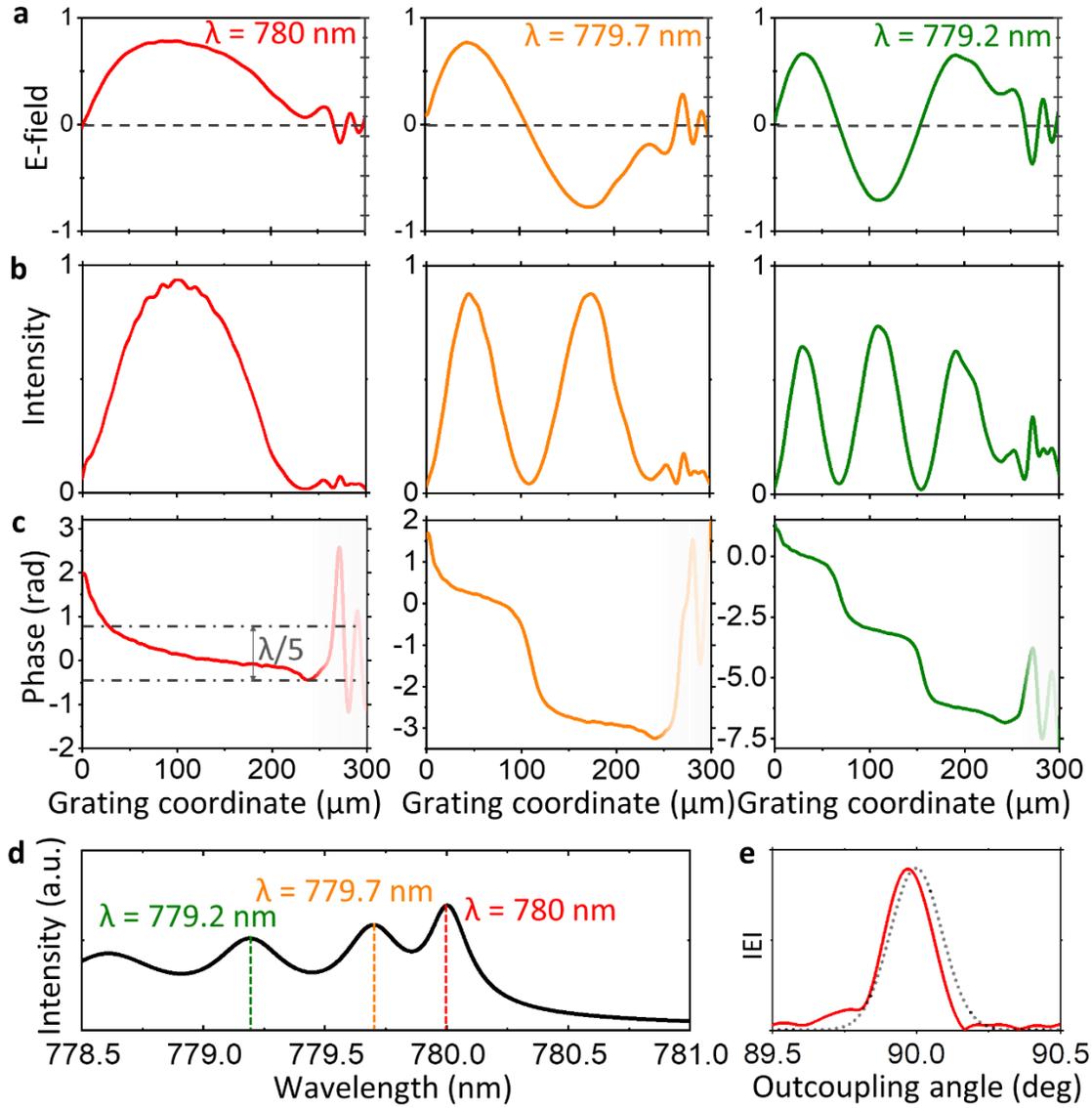

**Fig. 2 Simulated resonant grating performance. a, b** and **c** depict electric field, intensity profiles, and phase of the first three spatially broad resonant modes supported in the optimized grating and projected in free space at λ = 780 nm, 779.7 nm, and 779.2 nm. **d**, Outcoupled free-space power from the grating vs. laser wavelength. **e**, Far-field distribution of the fundamental mode vs. outcoupling angle (solid curve). The dotted curve depicts the Fourier transformed Gaussian fit of the fundamental mode assuming a perfectly collimated surface normal beam projection.

*Resonant grating performance*

The experimental far-field intensity profiles of the projected free-space beam depict the single-peak (fundamental) mode at 780.15 nm ± 0.03 nm wavelength (Figures 3 c,g) and other blue-shifted multiple-peak patterns (Figures 3 a,b,e,f). As expected, red-wavelength detuning falls into a band gap (Figures 3 d,h). The discrepancy between experimental and simulated data is attributed to fabrication imperfections. Since the experimental grating parameters such as slab thickness, grating DC, depth, and period define the accuracy of a resonant mode, by adjusting the fabrication process we can tune the device operating wavelength.

The experimental free-space intensity profiles shown in Figures 3 e-h correspond to the integrated fundamental, 2nd, 3rd resonant modes, and the intensity profile in the band gap (Figures 3 a-d) projected from the resonant grating at 780.15 nm ± 0.03 nm, 779.85 nm ± 0.03 nm, 779.35 nm ± 0.03 nm, and 780.35 nm ± 0.03 nm wavelengths, respectively. The experimental data qualitatively agree with simulated intensities plotted in Figure 2 b. The measured wavelength dependence of the total power transmitted to free space confirms that the fundamental mode (Figure 3 i) has the maximal transmission as a result of grating optimization. The Gaussian fit of the fundamental peak reveals ≈ 90 µm FWHM (Figure 3 g). The deviation from the designed value (≈ 118 µm FWHM) is due to fabrication imperfections, such as possibly slightly deeper or wider grating groves. Hence the mode size conversion ratio between the single-mode waveguide with cross-sectional dimensions of 250 nm × 300 nm and the projected Gaussian beam of 90 µm × 172 µm FWHM Gaussian beam is ≈ $2.1 \times 10^5$. The measured conversion efficiency of the fundamental resonance, defined as the total power from the grating measured in free-space relative to the waveguide power at the device, is ≈ 4.5 dB.

To quantify the free-space beam propagation, we acquire images of the outcoupled radiation (Figure 4 a) in focal planes located at known distances within a few millimeters above and below the chip surface (Figure 4 b). The beam's waist analysis based on the Gaussian fit for each image indicates that the radiation profile maintains its shape across at least 14 mm indicating good collimation with minimal wavefront distortion (Figure 4 c). The weak focusing observed above the chip surface is indicative of a small wave front curvature, which is attributed to fabrication imperfections. By quantifying the lateral beam shift in the direction across the grating lines as a function of the focal plane height, we find that the beam emanates close to a surface normal at an average angle of $0.06^o \, ^{+0.22^o}_{-0.07^o}$. The reported uncertainties correspond to the minimum and maximum bounds, obtained by fitting the measured data within different *z*-coordinate subdomains in Figure 4 c.

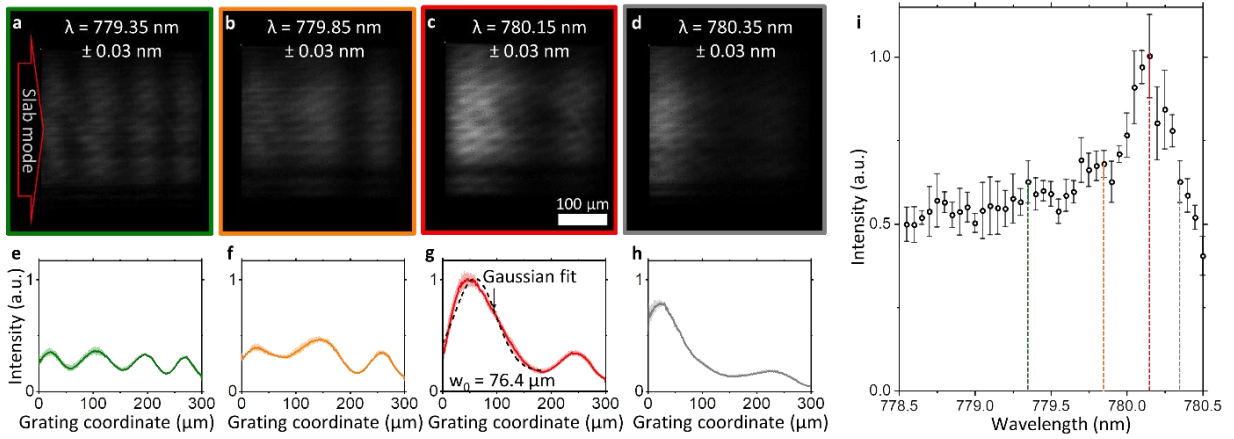

**Fig. 3 Experimental spectroscopic characterization of the resonant gratings. a-d**, Top-view optical images of the free-space beam projected from the photonic grating at 779.35 nm ± 0.03 nm, 779.85 nm ± 0.03 nm, 780.15 nm ± 0.03 nm, and 780.35 nm ± 0.03 nm wavelength depicting the 3rd, 2nd and fundamental resonant modes, and the intensity profile detuned into the band gap, respectively. Wavelength uncertainties are one standard deviation of the multiple calibrated lasing wavelength measurements across the range from 778.3 nm to 781.3 nm. **e-h,** Intensity profiles corresponding to images depicted in panels a-d. Dotted line in panel g is a Gaussian fit of the fundamental mode. **i,** Outcoupled free-space power from the grating, normalized to its maximum, vs. laser wavelength. Error bars are single standard deviations of the data obtained by equally dividing the images into 5 subdomains along the grating width.

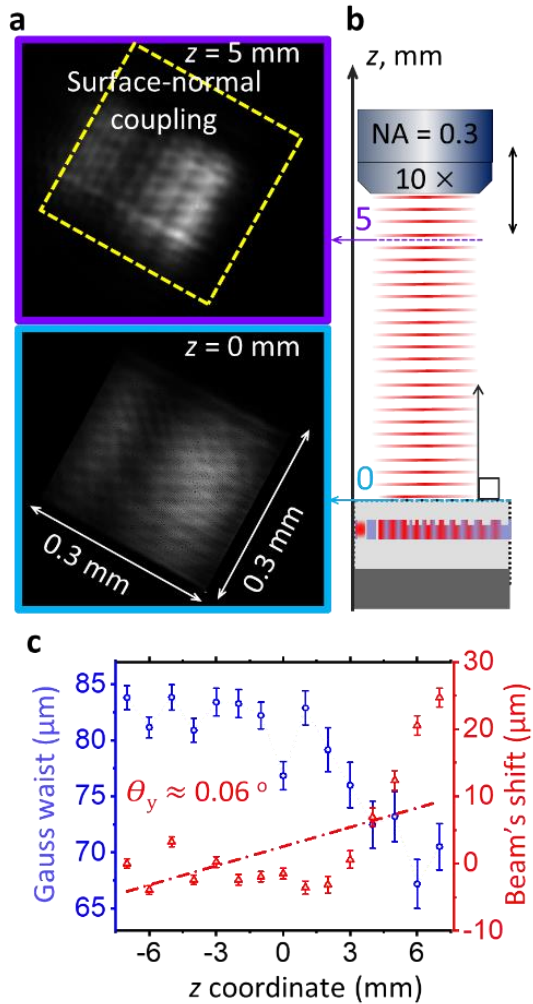

**Fig. 4 Characterization of the surface-normal free-space beam emanated from the grating coupler. a,** Free-space beam images acquired at focal planes at the chip surface (bottom image) and $z = 5$ mm above the chip (top image). The grating is tuned into the fundamental mode ($\lambda = 780.15$ nm ± 0.03 nm). Yellow dashed box points out the position of the resonant grating. **b,** A schematic showing image acquisition. **c,** Evolution of the measured Gaussian waist and the lateral shift of the free-space Gaussian beam vs. $z$ coordinate. The outcoupling angle is estimated from the beam's shift as a function of the $z$ coordinate. Dash-dot line is a linear fit of the beam's shift. Error bars correspond to one standard deviation of the data obtained from 5 equal image subdomains along the grating width.

*Thermo-optical tuning*

Finally, we demonstrate how we leverage the thermo-optical effect to precisely tune the device to a desired operating wavelength. Due to its resonant nature, the optimal emission wavelength of this device is sensitive to fabrication imperfections and may deviate from the desired value. Thermo-optical tuning provides means to compensate for fabrication deficiencies. Figure 5 illustrates how the free-space intensity profiles at a single wavelength vary with increasing device temperature within the range of ≈ 80 K from the room temperature. Initially tuned into the fundamental mode at the room temperature, the free-space beam intensity pattern transitions to high-order modes upon heating. The conducted experiments confirm that the device tunability is fully reversible and repeatable. By comparing the intensity profiles, the temperature variation can be equated to an equivalent wavelength change, as shown in Figure 5; e.g., heating the resonant grating by ≈ 80 K results in the same intensity profile change as tuning the wavelength by $\Delta\lambda \approx 1$ nm. Linear fit to the data in Figure 5 indicates the wavelength tuning slope of 0.012 nm/K ± 0.001 nm/K, with the one standard deviation uncertainty obtained from the fit parameter estimated variance.

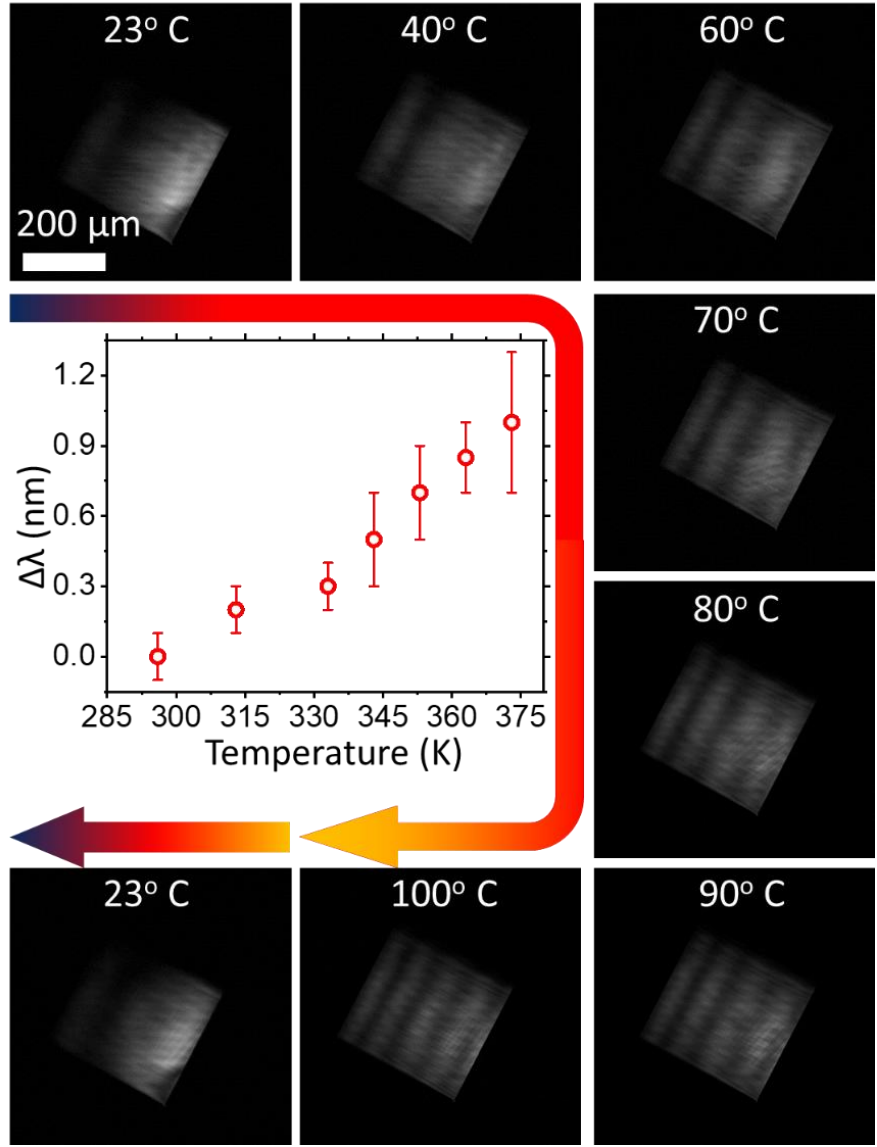

**Fig. 5 Thermo-optical tuning.** Flowchart depicts the sequence of optical images of the projected free-space beam upon heating the resonant grating about 80 K above room temperature and then cooling back to room temperature. Δλ represents the estimated optimum wavelength tuning upon sample heating. Error bars represent uncertainties in wavelength detuning between two neighboring images for which the outcoupled free-space beam profile can be uniquely correlated to a specific temperature pattern.

DISCUSSION

To summarize, we have demonstrated the deformation-based inverse design algorithm to optimize large photonic structures. The 300 μm long grating projects a surface-normal 90 μm FWHM collimated free-space beam from a photonic waveguide with high conversion efficiency. The unique performance of the novel photonic configuration is enabled due to spatially broad standing wave resonances, ensuring the surface-normal beam coupling due to an NZI operation. The slow-light grating modes ensure both the full mode matching minimizing reflection losses and free-space light intensity control. The experimentally realized photonic device is readily adaptable to operate at visible, telecom, and ultraviolet wavelengths and will further matchmake integrated photonics with free-space optics for novel chip-scale hybrid systems in science and technology realms with the immense need for planar 2D integration.

MATERIALS AND METHODS

*Fabrication of the photonic chip*

The resonant gratings are realized using the standard nanofabrication techniques (See SI for details). Nominally 2.9 μm thick $SiO_2$ and 250 nm $SiN_x$ continuous layers are formed using thermal oxidation and low-pressure chemical vapor deposition (LPCVD), respectively. Then electron-beam lithography and reactive-ion etching are employed twice to pattern fully etched high-confinement waveguides and evanescent expanders and 30 nm deep partially etched grating in the $SiN_x$ slab. The top ≈ 3 μm thick $SiO_2$ cladding is deposited using LPCVD as well. All grating dimensions are thoroughly characterized using scanning electron microscopy.

*Characterization of the device's operation*

We utilize a fiber-coupled laser source tunable around 780 nm wavelength to feed the photonic chip and an optical microscope equipped with a 0.3 NA objective backed with a complementary metal oxide semiconductor (CMOS) camera to characterize the free-space beam profiles, respectively. Before conducting experiments, we have calibrated the lasing wavelength by coupling to a commercial rubidium cell as a reference and also using a wavemeter. The characteristic Rb transitions occur close to the grating fundamental mode at 780 nm. We then tune the laser from the calibrated Rb transitions to the wavelength of the resonant grating fundamental mode using a high-repeatability scan stepper motor and check the lasing wavelength by a wavemeter. From calibration data, we estimate the wavelength uncertainty in our experiments due to mode hopping to be within ± 0.03 nm range, which represents one standard deviation statistical uncertainty of the calibrated lasing wavelength across the range from 778.3 nm to 781.3 nm. The laser radiation is coupled to a $TE_0$ waveguide single-mode via an inverted-taper. We use a fiber polarizer and an attenuator to adjust the polarization and power ($\approx$ 1 mW) of the input light, respectively. To characterize the direction of the free-space beam, we analyze optical images collected at the grating surface and a few millimeters above the chip.

DATA AVAILABILITY

The data produced in this study are available from the corresponding author upon reasonable request.

ACKNOWLEDGEMENTS

Dr. Alexander Yulaev acknowledges support under the Professional Research Experience Program (PREP), administered through the Department of Chemistry and Biochemistry, UMD.

AUTHOR CONTRIBUTIONS

VAA conceived the idea and directed the work. DAW fabricated samples. AY characterized samples. AY and VAA contributed to the data interpretation. The manuscript was written through contributions of all authors. All authors have given approval to the final version of the manuscript.

CONFLICT OF INTEREST

The authors declare that they have no conflict of interest.

Correspondence and requests for materials should be addressed to V. A. A. (email: *vladimir.aksyuk@nist.gov*)

REFERENCES

1    Kippenberg, T. J., Holzwarth, R. & Diddams, S. A. Microresonator-based optical frequency combs. *Science* **332**, 555-559 (2011).
2    Kitching, J. *et al.* in *Journal of Physics: Conference Series.*  012056 (IOP Publishing).
3    Grier, D. G. A revolution in optical manipulation. *Nature* **424**, 810 (2003).
4    Hummon, M. T. *et al.* Photonic chip for laser stabilization to an atomic vapor with 10−11 instability. *Optica* **5**, 443-449 (2018).
5    Kohnen, M. *et al.* An array of integrated atom–photon junctions. *Nature Photonics* **5**, 35-38 (2011).
6    Mehta, K. K. *et al.* Integrated optical addressing of an ion qubit. *Nature Nanotechnology* **11**, 1066 (2016).
7    Lin, S. & Crozier, K. B. Trapping-assisted sensing of particles and proteins using on-chip optical microcavities. *ACS Nano* **7**, 1725-1730 (2013).
8    Xu, D.-X. *et al.* Folded cavity SOI microring sensors for high sensitivity and real time measurement of biomolecular binding. *Optics Express* **16**, 15137-15148 (2008).
9    Sun, Y. & Fan, X. Optical ring resonators for biochemical and chemical sensing. *Analytical and Bioanalytical Chemistry* **399**, 205-211 (2011).
10   Squires, T. M., Messinger, R. J. & Manalis, S. R. Making it stick: convection, reaction and diffusion in surface-based biosensors. *Nature Biotechnology* **26**, 417-426 (2008).
11   Dantham, V., Holler, S., Kolchenko, V., Wan, Z. & Arnold, S. Taking whispering gallery-mode single virus detection and sizing to the limit. *Applied Physics Letters* **101**, 043704 (2012).
12   Zhu, J. *et al.* On-chip single nanoparticle detection and sizing by mode splitting in an ultrahigh-Q microresonator. *Nature Photonics* **4**, 46 (2010).
13   Thomson, D. *et al.* Roadmap on silicon photonics. *Journal of Optics* **18**, 073003 (2016).
14   Dakss, M., Kuhn, L., Heidrich, P. & Scott, B. Grating coupler for efficient excitation of optical guided waves in thin films. *Applied Physics Letters* **16**, 523-525 (1970).
15   Harris, J., Winn, R. & Dalgoutte, D. Theory and design of periodic couplers. *Applied Optics* **11**, 2234-2241 (1972).
16   Vedantam, S. *et al.* A plasmonic dimple lens for nanoscale focusing of light. *Nano Letters* **9**, 3447-3452 (2009).
17   Yariv, A. & Yeh, P. *Photonics: optical electronics in modern communications (the oxford series in electrical and computer engineering)*.  (Oxford University Press, Inc., 2006).
18   Ogawa, K., Chang, W., Sopori, B. & Rosenbaum, F. A theoretical analysis of etched grating couplers for integrated optics. *IEEE Journal of Quantum Electronics* **9**, 29-42 (1973).
19   Taillaert, D., Bienstman, P. & Baets, R. Compact efficient broadband grating coupler for silicon-on-insulator waveguides. *Optics Letters* **29**, 2749-2751 (2004).
20   Van Laere, F. *et al.* Compact and highly efficient grating couplers between optical fiber and nanophotonic waveguides. *Journal of Lightwave Technology* **25**, 151-156 (2007).
21   Ding, Y., Peucheret, C., Ou, H. & Yvind, K. Fully etched apodized grating coupler on the SOI platform with− 0.58 dB coupling efficiency. *Optics Letters* **39**, 5348-5350 (2014).

22  Roelkens, G. *et al.* High efficiency diffractive grating couplers for interfacing a single mode optical fiber with a nanophotonic silicon-on-insulator waveguide circuit. *Applied Physics Letters* **92**, 131101 (2008).
23  Vermeulen, D. *et al.* High-efficiency fiber-to-chip grating couplers realized using an advanced CMOS-compatible silicon-on-insulator platform. *Optics Express* **18**, 18278-18283 (2010).
24  Ding, Y., Ou, H. & Peucheret, C. Ultrahigh-efficiency apodized grating coupler using fully etched photonic crystals. *Optics Letters* **38**, 2732-2734 (2013).
25  Zaoui, W. S. *et al.* Bridging the gap between optical fibers and silicon photonic integrated circuits. *Optics Express* **22**, 1277-1286 (2014).
26  Chen, X., Li, C., Fung, C. K., Lo, S. M. & Tsang, H. K. Apodized waveguide grating couplers for efficient coupling to optical fibers. *IEEE Photonics Technology Letters* **22**, 1156-1158 (2010).
27  Mehta, K. K. & Ram, R. J. Precise and diffraction-limited waveguide-to-free-space focusing gratings. *Scientific Reports* **7**, 1-8 (2017).
28  Mekis, A. *et al.* A grating-coupler-enabled CMOS photonics platform. *IEEE Journal of Selected Topics in Quantum Electronics* **17**, 597-608 (2010).
29  Xu, X. *et al.* Complementary metal–oxide–semiconductor compatible high efficiency subwavelength grating couplers for silicon integrated photonics. *Applied Physics Letters* **101**, 031109 (2012).
30  Song, J. H. *et al.* Polarization-independent nonuniform grating couplers on silicon-on-insulator. *Optics Letters* **40**, 3941-3944 (2015).
31  Chen, X., Xu, K., Cheng, Z., Fung, C. K. & Tsang, H. K. Wideband subwavelength gratings for coupling between silicon-on-insulator waveguides and optical fibers. *Optics Letters* **37**, 3483-3485 (2012).
32  Halir, R. *et al.* Continuously apodized fiber-to-chip surface grating coupler with refractive index engineered subwavelength structure. *Optics Letters* **35**, 3243-3245 (2010).
33  Kim, S. *et al.* Photonic waveguide to free-space Gaussian beam extreme mode converter. *Light: Science & Applications* **7**, 72 (2018).
34  Yulaev, A. *et al.* Metasurface-Integrated Photonic Platform for Versatile Free-Space Beam Projection with Polarization Control. *ACS Photonics* **6**, 2902-2909 (2019).
35  Yulaev, A. *et al.* in *CLEO: Applications and Technology.*  AM3K. 2 (Optical Society of America).
36  Chauhan, N. *et al.* in *2019 Conference on Lasers and Electro-Optics (CLEO).*  1-2 (IEEE).
37  Badham, K. *et al.* in *2017 Conference on Lasers and Electro-Optics Pacific Rim (CLEO-PR).*  1-5 (IEEE).
38  Baba, T. Slow light in photonic crystals. *Nature Photonics* **2**, 465-473 (2008).
39  Schulz, S. *et al.* Dispersion engineered slow light in photonic crystals: a comparison. *Journal of Optics* **12**, 104004 (2010).
40  Notomi, M. *et al.* Extremely large group-velocity dispersion of line-defect waveguides in photonic crystal slabs. *Physical Review Letters* **87**, 253902 (2001).
41  Magno, G. *et al.* Strong coupling and vortexes assisted slow light in plasmonic chain-SOI waveguide systems. *Scientific Reports* **7**, 1-11 (2017).
42  Gan, Q., Fu, Z., Ding, Y. J. & Bartoli, F. J. Ultrawide-bandwidth slow-light system based on THz plasmonic graded metallic grating structures. *Physical Review Letters* **100**, 256803 (2008).

43  Huang, Y., Min, C. & Veronis, G. Subwavelength slow-light waveguides based on a plasmonic analogue of electromagnetically induced transparency. *Applied Physics Letters* **99**, 143117 (2011).
44  Reshef, O. *et al.* Direct observation of phase-free propagation in a silicon waveguide. *ACS Photonics* **4**, 2385-2389 (2017).
45  Li, Y. *et al.* On-chip zero-index metamaterials. *Nature Photonics* **9**, 738-742 (2015).
46  Liberal, I. & Engheta, N. Near-zero refractive index photonics. *Nature Photonics* **11**, 149-158 (2017).
47  Kinsey, N., DeVault, C., Boltasseva, A. & Shalaev, V. M. Near-zero-index materials for photonics. *Nature Reviews Materials*, 1-19 (2019).
48  Yulaev, A. *et al.* in *2018 International Conference on Optical MEMS and Nanophotonics (OMN).* 1-2 (IEEE).
49  Molesky, S. *et al.* Inverse design in nanophotonics. *Nature Photonics* **12**, 659-670 (2018).
50  Piggott, A. Y. *et al.* Inverse design and demonstration of a compact and broadband on-chip wavelength demultiplexer. *Nature Photonics* **9**, 374-377 (2015).
51  Michaels, A. & Yablonovitch, E. Inverse design of near unity efficiency perfectly vertical grating couplers. *Optics Express* **26**, 4766-4779 (2018).

Supplementary information for

# Surface-Normal Free-Space Beam Projection via Slow-Light Standing Wave Resonances in Extra-Large Near-Zero Index Grating Couplers

*Alexander Yulaev[1,2], Daron A. Westly[1], and Vladimir A. Aksyuk[1]**

*[1]Physical Measurement Laboratory, National Institute of Standards and Technology, Gaithersburg, MD 20899, USA.*

*[2]Department of Chemistry and Biochemistry, University of Maryland, College Park, MD 20742, USA.*

**Corresponding author: vladimir.aksyuk@nist.gov*

**Far-field simulation of free-space beams outcoupled from standing-wave resonances using finite element method (FEM)**

Simulated performance in Figure S1 reveals that the resonant grating is optimized to couple forward grating mode to the counterpropagating grating mode to obtain a standing wave resonance.

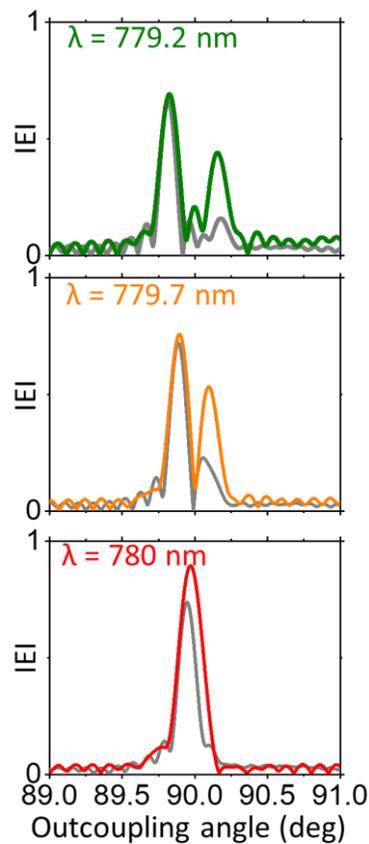

Figure S1. Far-field FEM simulations of the free-space radiation outcoupled from the first 3 standing-wave resonances: the fundamental resonance (red curve), the second mode (yellow curve), and the third mode (green curve). Gray curves depict the far-field radiation from the corresponding uniform gratings.

# Fabrication of resonant gratings

2.9 μm thick wet thermal oxide is grown on Si followed by 250 nm thick $SiN_x$ deposition using LPCVD. The $SiN_x$ is pattered twice using EBL and RIE to define shallow gratings, waveguide, and evanescent expander structures. Nominally 3 μm thick cladding is deposited using LPCVD. The wafer is diced in chips, and chip edges are polished to couple light from fiber to a $SiN_x$ waveguide.

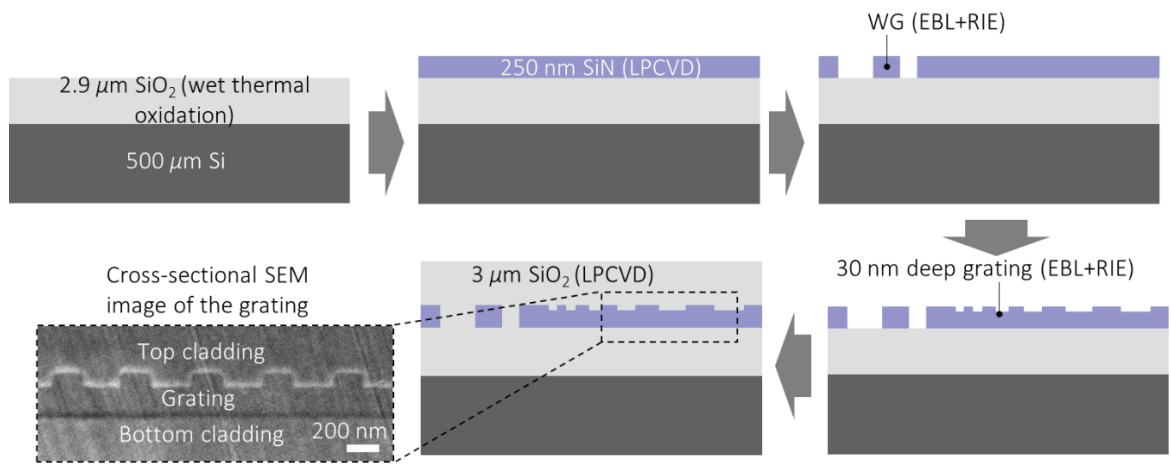

Figure S2. Fabrication of resonant gratings. LPCVD - low-pressure chemical vapor deposition. WG – a waveguide. EBL - electron-beam lithography. RIE - reactive-ion etching.

# Experimental set-up

A fiber-coupled laser is used to excite standing wave resonances. The power and polarization of the input light are adjusted using a fiber attenuator and a polarizer, respectively. An optical microscope equipped with 0.3 NA objective and backed with CMOS camera are employed to characterize device performance.

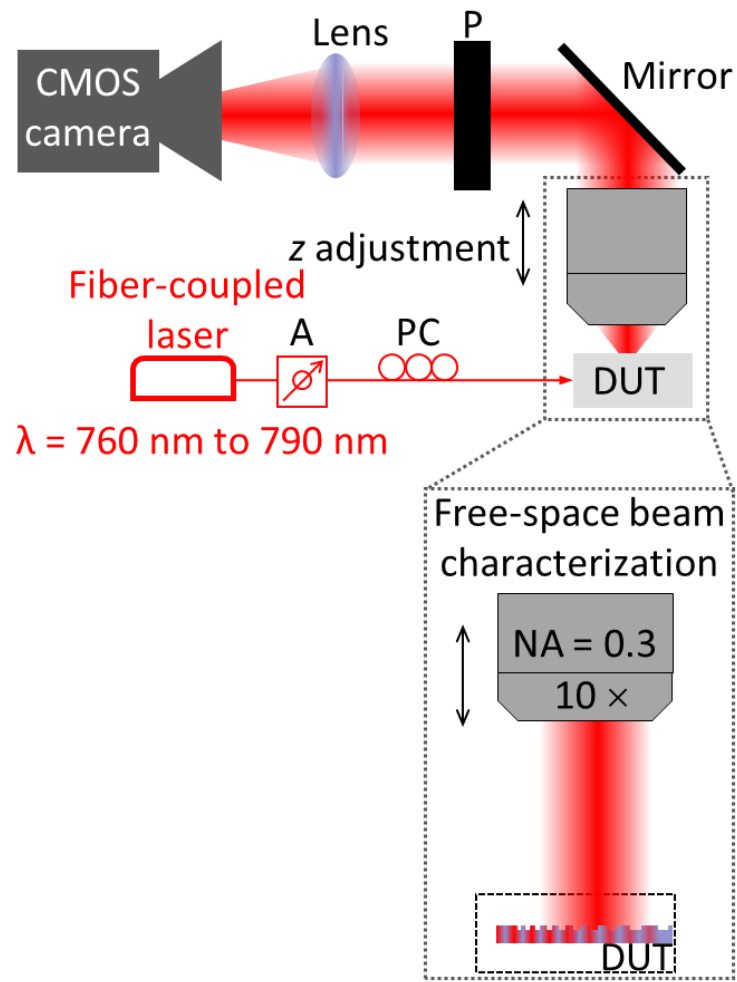

Figure S3. Experimental set-up. P – free-space polarizer. PC – fiber polarizer. A – fiber attenuator. DUT – device under test.

**Free-space radiation intensity map *versus* laser wavelength**

The experimental integrated map of free-space beam radiation as a function of grating coordinate and laser wavelength is obtained by integrating optical images over a grating line direction.

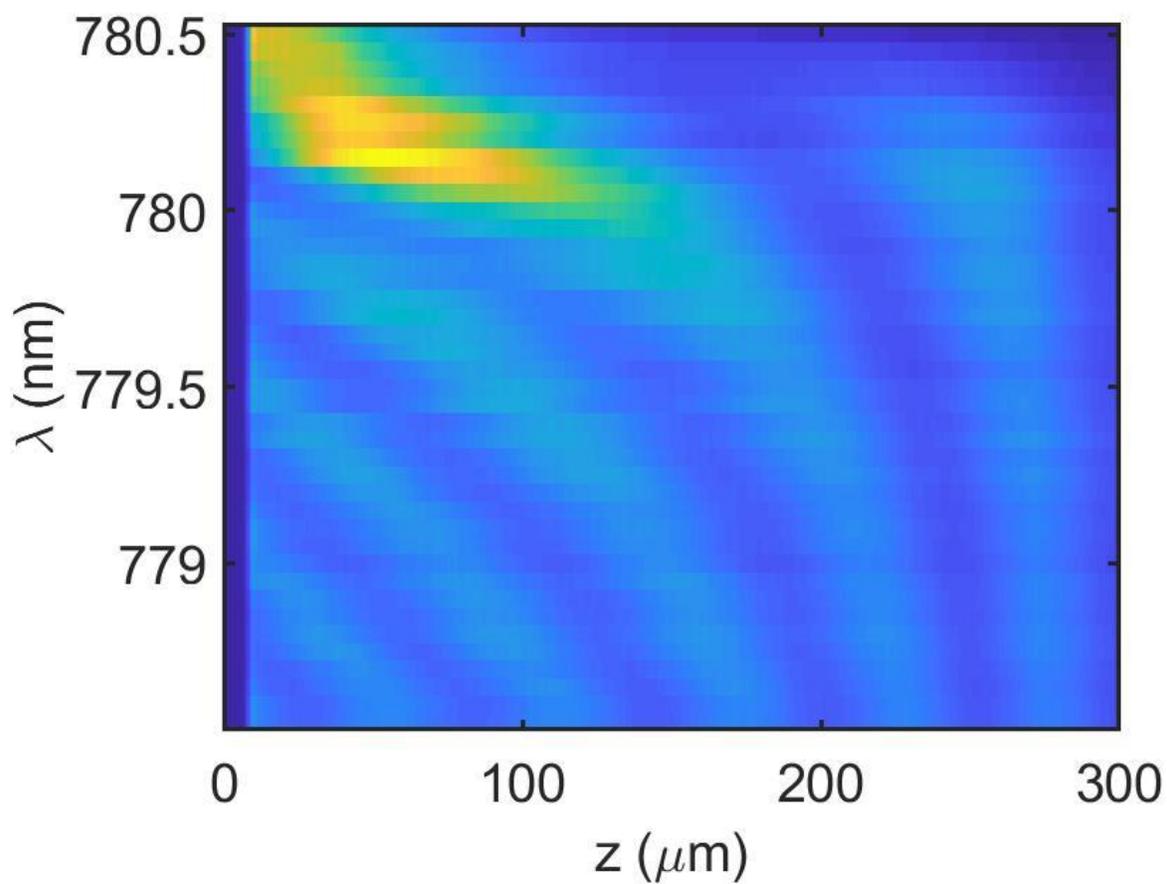

Figure S4. Experimental spectral dependence of free-space radiation intensity

**Thermo-optical tuning**

(i)  *Simulations*

Simulations of resonant grating performance at different temperatures are calculated using FEM where dependence of refractive material indices on temperature is tabulated.

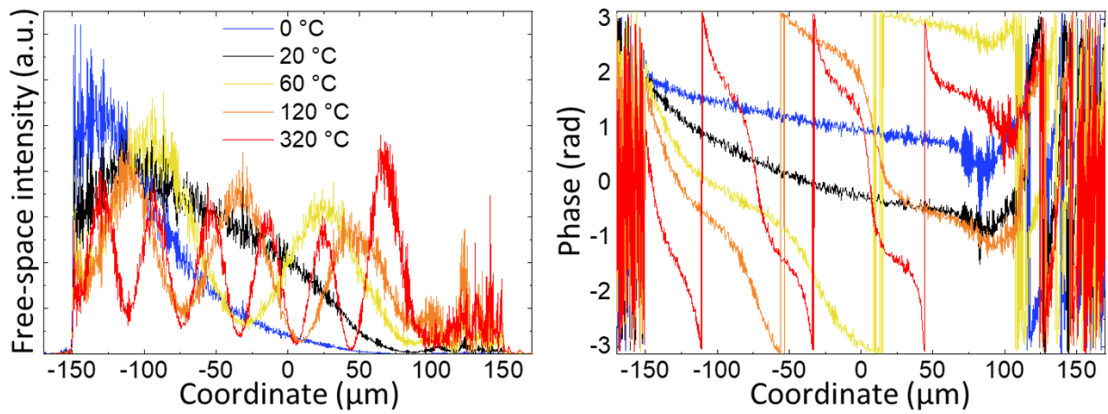

Materials refractive indices vs. temperature

| Temp, deg C | SiO | SiN | Si |
|---|---|---|---|
| 0 | 1.399828 | 1.99952 | 3.3957 |
| 20 | 1.4 | 2 | 3.4 |
| 60 | 1.400344 | 2.00096 | 3.4086 |
| 120 | 1.40086 | 2.0024 | 3.4215 |
| 320 | 1.40258 | 2.0072 | 3.4645 |

Arbabi, Amir, and Lynford L. Goddard, Optics letters **38**, p. 3878 (2013)

Figure S5. Simulated resonance grating performance during thermo-optical tuning. Top plots depict free-space intensity profiles (left), and the phase (right) along the grating coordinate upon heating the device up to 0 °C, 20 °C, 60 °C, 120 °C, and 320 °C. The grating performance is simulated using FEM with temperature-dependent material refractive indices specified in the table.

*(ii)    Experiment*

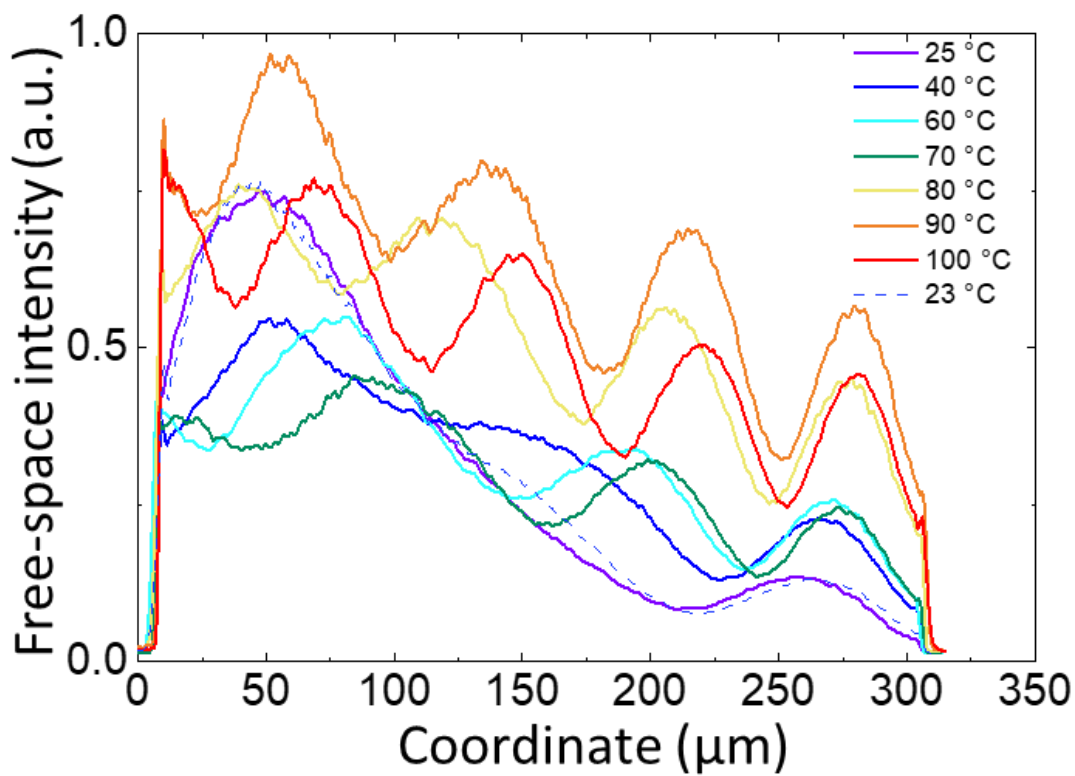

Figure S6. The experimental characterization of thermo-optical tuning is conducted by controlling the sample temperature using a thermal controller. The sample temperature increases from 25 ºC to 100 ºC while the outcoupled free-space beam is routinely characterized. Then the sample is cooled down to 23 ºC to test reproducibility of thermo-optical tuning.